\documentclass{elsart}

\newcommand{\beq}{\begin{equation}}
\newcommand{\eneq}{\end{equation}}

\usepackage{amssymb,graphicx}

\begin{document}

\begin{frontmatter}



\title{ Effective Boundary Field Theory for a Josephson  Junction
Chain with a Weak Link}


\author{Domenico Giuliano $^{\dagger }$ and
 Pasquale Sodano $^{ \ast }$}

\address{$^{\dagger }$ Dipartimento di Fisica, Universit\`{a} della
Calabria and  I.N.F.N., Gruppo collegato \\
di Cosenza, Arcavacata di Rende I-87036, Cosenza, Italy  \\
(E-Mail: {\rm giuliano@fis.unical.it}) \\ $^\ast$
Dipartimento di Fisica e Sezione I.N.F.N.,
 Universit\`a di Perugia,\\  Via A. Pascoli I-06123, Perugia, Italy \\
(E-Mail: {\rm pasquale.sodano@pg.infn.it})}

\begin{abstract}
We show that  a finite Josephson Junction (JJ) chain, ending with
two bulk superconductors, and  with a weak link at its center, may
be regarded as a condensed matter realization of a two-boundary
Sine-Gordon model. Computing the partition function yields a
remarkable analytic expression for the DC Josephson current as a
function of the phase difference across the chain. We show that, in
a suitable range of the chain parameters, there is a crossover of
the DC Josephson current from a sinusoidal to a sawtooth behavior,
which signals a transition from a regime where the boundary term is
an irrelevant operator to a regime where it becomes relevant.
\end{abstract}

\end{frontmatter}


\section{Introduction}

In this paper, we analyze a superconducting 1+1-dimensional system,
defined on a finite interval of length $L$. If  the bulk is
described by a massless theory, and conformal boundary conditions
are chosen, one could understand the properties of the model, using
the formalism of boundary conformal field theories \cite{cardy}. If
one deviates from this situation by either adding an interaction in
the bulk, or at the boundary, or both, the behavior of the system
becomes much more interesting, since it involves crossovers
depending on the bulk and boundary energy scales, as well as on the
finite size $L$.

In the sequel, we shall show that a JJ-chain with a weak link at its
center and ending with two bulk superconductors at fixed phase
difference $\varphi$, is the prototype of a condensed matter
realization of a two-boundary Sine-Gordon model \cite{2bsg}, whose
Hamiltonian is given by

\beq
H = \frac{1}{4 \pi} \int_0^L d x \: \left[ \frac{1}{v} \left( \frac{
\partial \Phi}{ \partial t} \right)^2 + v \left( \frac{
\partial \Phi}{ \partial x} \right)^2 \right]
- \Delta_L \cos \left[ \frac{\sqrt{g}}{2} \Phi ( 0 ) \right]
- \Delta_R \cos \left[  \frac{ \sqrt{g} \Phi ( L  ) - \varphi  }{2}
\right] \:\:\: .
\label{eq1.1}
\eneq
\noindent

Boundary field theories appear to be relevant in several different
contexts. In condensed matter physics, they are mostly
generalizations of quantum impurity models, which may be described
by using the Tomonaga-Luttinger Liquid (TLL)-paradigm
\cite{luttinger}; for instance, boundary interactions appear in the
analysis of the Kondo problem \cite{iaff}, in the study of a
one-dimensional conductor in presence of an impurity
\cite{impurity}, and in the derivation of the tunneling between edge
states of a Hall bar \cite{ludwighall}. The TLL paradigm shows that
many interactions are simply diagonalizable in the basis of
appropriate collective bosonic modes, and that non diagonalizable
interaction usually correspond to exactly solvable Hamiltonians,
such as Sine-Gordon models \cite{umklapp,exact}. Recently, boundary
field theories have been investigated in the context of string
theories. For instance, in studying tachyon instabilities
\cite{tachyon}, one is faced with the fact that the space of
interacting string theory \cite{sen} may be mapped onto the space of
boundary perturbations of conformal theories \cite{leclair}, and
that the renormalization group flow determined by boundary
perturbations may be identified with tachyon condensation
\cite{gava}. Affleck and Ludwig \cite{gthe} showed that the boundary
entropy $g$ is decreasing along the renormalization group
trajectories, triggered by the boundary interaction.

In a inspiring paper \cite{glar},  Glazman and Larkin analyzed the
quantum phase diagram of a  JJ-chain in the $V_g-J$-plane, where
$V_g$ is an external gate voltage applied to each junction, while
$J$ is the Josephson coupling between neighboring grains. They found
evidence that this system undergoes a phase transition between an
attractive TLL phase, with $g < 1$, and a repulsive TLL phase, with
$g > 1$. While the former phase is the one-dimensional analog of the
superconducting phase \cite{seb}, the repulsive Tomonaga-Luttinger
phase is peculiar of a one-dimensional system \cite{luttinger}. To
be self-contained, here we shall provide a detailed
field-theoretical description of the one-dimensional infinite chain
analyzed in Ref.\cite{glar}: our rather pedagogical derivation
evidences how the one-dimensional JJ-chain may be described in terms
of interacting 1+1-dimensional chiral fermions and how, using the
TLL paradigm, the interaction is exactly diagonalized in a pertinent
basis. The TLL-$g$ parameter \cite{shultz} is crucial for the
analysis of the phase diagram. Indeed, while for $g<1$ the system
supports an attractive TLL phase (superconducting), for $g>1$  the
JJ-chain is described by a repulsive TLL phase \cite{glar}. The
$g=1$-line corresponds to a noninteracting TLL model. All this
features may be quantitavely derived within the framework of the
bosonized 1+1-dimensional TLL-model\footnote{Notice that here the
TLL-g parameter is the inverse of the parameter used in Ref.\cite{glar}.}.

Using the TLL paradigm, we show that a finite JJ-chain with a weak
link at its center is mapped onto a two-boundary Sine-Gordon model,
with fixed Dirichlet boundary conditions at the outer boundary, and
with dynamical boundary conditions at the inner boundary. To study
the effects of the interaction at the inner boundary, we perform a
Renormalization Group (RG)  analysis, to derive how the effective
parameters of the system scale as a function of the size of the
chain $L$. We find that in the repulsive TLL-phase ($g>1$) the
boundary term is perturbative for any size $L$. At variance, in the
attractive TLL-phase ($g<1$), we find that there is an RG-invariant
length, $L^*$, such that, for $L < L^*$ the boundary term is
perturbative, while for $L \geq L^*$ it becomes nonperturbative. As
for the models analyzed in \cite{2bsg}, the crossover from the
perturbative to the nonperturbative regime is evidenced by a change
of the DC Josephson current  (as a function of the phase difference
at the bulk superconductors $\varphi$) from a sinusoidal to a
sawtooth behavior.

The paper is organized as follows:

\begin{itemize}

\item In Section II we analyze the infinite one-dimensional JJ-array described
in Ref.\cite{glar}
and provide a  detailed derivation of the mapping of this chain onto
the anisotropic ($XXZ$) spin 1/2 model;

\item In Section III we construct the effective field theory for the
equivalent $XXZ$ chain. We bosonize the theory and identify the various
parameters of the continuum model in terms of the microscopic parameters
of the lattice Hamiltonian;

\item In Section IV, using the TLL-paradigm, we derive the phase diagram
of the JJ-chain;

\item In Section V we show that the effective field theory for the
JJ-chain with
a weak link and ending with two bulk superconductors is indeed the
two-boundary Sine-Gordon model;

\item In Section VI, using the Coulomb Gas Renormalization
Group approach, we provide a careful estimate of the partition
function of the two-boundary Sine-Gordon model for any value of $g$.
We then derive the DC Josephson current across the chain, as a
function of $\varphi$, at both fixed points and explicitly show the
existence of a crossover from a sinusoidal to a sawtooth behavior;

\item Section VII is devoted to a discussion of our results.

\end{itemize}

\section{Mapping of the one-dimensional JJ-chain onto the $XXZ$ spin-1/2
model}

The simplest model Hamiltonian describing a one-dimensional JJ-chain is
given by:

\beq H = H_C + H_J \equiv \frac{E_C}{2} \sum_{j=1}^{L/a} \left( - i
\frac{\partial}{\partial \phi_j} - \frac{ {\it N}}{2} \right)^2 -  J
\sum_{j=1}^{L/a} \cos ( \phi_j - \phi_{j + 1 } ) \:\:\: .
\label{equ2.1} \eneq \noindent In Eq.(\ref{equ2.1}) $ - i
\frac{\partial}{\partial \phi_j}$ is  the operator representing the
number of Cooper pairs at site $j$ in the phase representation and,
thus,  it takes only integer eigenvalues, $n_j$,  $E_C$ is the
charging energy of a grain, $J$ is the Josephson coupling energy and
${\it N}$ accounts for the influence of a gate voltage, since $e
{\it N} \propto V_g$. The sum over $j$ ranges over the ($L/a$)
sites, with  $L$ being the length of the chain, and $a$ being the
intergrain distance; imposing periodic boundary conditions amounts
to fix $\phi_{L / a  + j } = \phi_j$. For $J / E_C \to 0$, the chain
is an insulator for almost any value of ${\it N}$, since it costs an
energy $\sim E_C$, to change  the number of pairs at any grain:
$E_C$ measures, then,  the insulating gap.

When ${\it N} = n + 1 / 2 + h$, with integer $n$ and $h \ll 1$, the two
states $n_j = n$ and $n_j = n + 1$ become almost degenerate in energy,
even for large $E_C$; in this limit one may restrict the
set of physical states to the Fock space ${\it F}$ spanned by the $2^{L/a}$
states

\[
| \{ n \} \rangle = \prod_{ j = 1}^{ L / a } | n_j \rangle \;\;\;\; ; \;
n_j = n , n + 1 \;\;\; .
\]
\noindent
The Josephson coupling lifts the degeneracy between $n$ and $n + 1$, since
$H_J$ may be represented as

\beq
H_J = - \frac{J}{2} [ e^{ i \phi_{j + 1} } e^{ - i \phi_j } +
e^{ i \phi_j } e^{ - i \phi_{j + 1 } } ] \;\;\; ,
\label{equ2.2}
\eneq
\noindent
with the operator  $e^{i \phi_j}\:\: (e^{-i \phi_j})$ raising (lowering)
the charge $n_j$ by 1.

Resorting to  the a well known procedure \cite{swolff}, one may
easily construct the effective Hamiltonian $H_{\rm eff}$, describing
the JJ-chain on the reduced space  ${\it F}$ \cite{glar}. Let $P$ be
the projector onto ${\it F}$ and $P_\perp$ be the projector onto the
subspace ${\it F}_\perp$, to ${\it O} (J^2 / E_C)$, $H_{\rm eff}$
takes the form:

\beq
H_{\rm eff}  = P ( H_J + H_C )  P + P \left[ \frac{H_J P_\perp H_J }{ -
\frac{9}{16} E_C } \right] P \:\:\: .
\label{equ2.3}
\eneq
\noindent

When restricted to ${\it F}$, the operators
$e^{\pm i \phi_j}$ and $ - i \frac{\partial}{\partial \phi_j}$  may be
represented with the spin-1/2 operators  $S_j^\pm$ and $S_j^z$, as

\beq
P e^{ \pm i \phi_j } P = S_j^\pm \;\;\; ; \;
P \left( - i \frac{\partial}{\partial \phi_j} - n - \frac{1}{2} \right) P =
S_j^z \:\:\: .
\label{equ2.4}
\eneq
\noindent
>From Eq.(\ref{equ2.4}), one immediately sees that a charge
$n$ ($n+1$)-state corresponds to a spin-(-1/2) (1/2)- eigenstate
of $S_j^z$, and that, to ${\it O} ( J^2/E_C)$, $H_{\rm eff}$ is given by

\beq
H_{\rm eff} = - \frac{J}{2}  \sum_{ j = 1}^{ L / a }  [ S_j^+
S_{j + 1}^- + S_{j + 1}^+ S_j^- ]
- E_C h \sum_{j = 1}^{ L / a } S_j^z
- \frac{3}{16} \frac{J^2}{E_C} \sum_{j = 1}^{L / a } S_j^z S_{j + 1}^z \:\:\: .
\label{equ2.5}
\eneq
\noindent

To account for the contributions coming from intergrain
capacitances, it is sufficient to retain only the nearest-neighbor
terms \cite{glar}, since  next-to-nearest neighbor hopping terms
would  give rise to irrelevant operators, and to add to the
Hamiltonian (\ref{equ2.5}) the term $E_Z \sum_{j = 1}^{L / a } S_j^z
S_{j + 1}^z $ ($E_Z > 0$) \cite{glar}. Thus, the system is usefully
described \cite{glar} by

\beq
H_{\rm Eff} =  - \frac{J}{2}  \sum_{ j = 1}^{ L / a }  [ S_j^+
S_{j + 1}^- + S_{j + 1}^+ S_j^- ]
 - H \sum_{j = 1}^{ L / a } S_j^z +
\Delta \sum_{j = 1}^{L / a } S_j^z S_{j + 1}^z
\label{equ2.6} \;\;\; ,
\eneq
\noindent
with $H = E_C h$ and $\Delta = E_Z - \frac{3}{16} \frac{J^2}{E_C}$.

Eq.(\ref{equ2.6}) is the Hamiltonian for a spin-1/2 $XXZ$-chain in
an external magnetic field $H$. The anisotropy parameter $\Delta$
may take positive, as well as negative values, depending on the
constructive parameters of the JJ-chain. As elucidated in the
following Sections, the sign of $\Delta$ is crucial for the
emergence of a Repulsive Tomonaga-Luttinger (RTL) phase in a
JJ-chain.

\section{Continuum field theory and bosonization of the $XXZ$-chain}

Using the standard bosonization technique \cite{shultz}, we derive
in this Section the effective low-energy long-wavelength field
theory associated to the Hamiltonian (\ref{equ2.6}). For this
purpose, one starts to write  the spin operators $S_j^a$ in terms of
Jordan-Wigner (JW) spinless lattice fermions $a_j$
\cite{jordanwigner},  obeying standard anticommutation relations:

\beq
\{ a_j , a_k^\dagger \} = \delta_{ j k } \:\:\: .
\label{equ3.1}
\eneq
\noindent

The JW transformation amounts to define:

\beq
S_j^+ \equiv a_j^\dagger \exp [ i \pi \sum_{ l = 1}^{ j - 1 }
a_l^\dagger a_l ]
\; \; \; ; \;
S_j^z \equiv a_j^\dagger a_j - \frac{1}{2}
\; \; \; ; \;
S_j^- \equiv  \exp [- i \pi \sum_{ l = 1}^{ j - 1 }
a_l^\dagger a_l ] a_j \:\:\: ,
\label{equ3.2}
\eneq
\noindent
which, in turn, implies:

\beq
[ S_j^a , S_i^b ] = i \delta_{j i } \epsilon^{a b c} S^c_j
\label{equ3.3} \:\:\: .
\eneq
\noindent
>From Eqs.(\ref{equ3.2}), the fermionic effective Hamiltonian
 may be written as:

\[
H_{\rm Eff}^f \equiv H_K + H_P +  H \sum_{ j = 1}^{L / a } a_j^\dagger a_j
=
\]

\beq
 - \frac{J}{2} \sum_{ j = 1}^{L / a} [ a_j^\dagger a_{j + 1}
+ a_{j + 1}^\dagger a_j ] +
\Delta \sum_{ j = 1}^{L / a}
( a_j^\dagger a_j
- \frac{1}{2} ) ( a_{ j + 1}^\dagger a_{ j + 1 } - \frac{1}{2} )
+ H \sum_{ j = 1}^{L / a } a_j^\dagger a_j \: .
\label{equ3.4}
\eneq
\noindent

The hopping term $H_K$ is readily diagonalized by resorting to the Fourier
components of $a_j$, $a_k$,

\beq
a_j = \frac{1}{ \sqrt{L / a }} \sum_k a_k e^{ i k ( j a )}
\label{equ3.5} \:\:\: ,
\eneq
\noindent
leading to:

\beq
H_K = \sum_k \epsilon ( k ) a_k^\dagger a_k \;\;\; ; \;\;
( \epsilon ( k ) = - J \cos ( k a ) - H )
\label{equ3.6}
\eneq
\noindent

If  $ | H | < J$, the Fermi surface is disconnected and consists
of two isolated points at $ \pm k_F$, with $ k_F = {\rm arccos} ( H / J )$.
Keeping only the excitations about the Fermi points with momenta $k$ such that
$ | k  \pm k_F | \leq \Lambda $, one obtains:

\[
H_K \approx \sum_{ | k - k_F | \leq \Lambda } \epsilon ( k ) a_k^\dagger a_k
+  \sum_{ | k + k_F | \leq \Lambda } \epsilon ( k ) a_k^\dagger a_k
\approx
J \sin ( a k_F )  \sum_{ | p | \leq \Lambda} \sin ( p a )
a_L^\dagger ( p ) a_L ( p )
\]

\beq
- J \sin ( a k_F )  \sum_{ | p | \leq \Lambda}
\sin ( p a ) a_R^\dagger ( p ) a_R ( p ) \:\:\: ,
\label{equ3.7}
\eneq
\noindent

with:

\[
a_L ( p ) \equiv a_{ p + k_F} \;  \;\; ; \;
a_R ( p ) \equiv a_{ p - k_F} \; ( | p | \le \Lambda ) \:\:\: .
\]
\noindent

For $\Lambda \ll k_F$, one may define the continuum chiral fields
$\psi_{L / R} ( x )$ as

\beq
\frac{ a_j }{ \sqrt{2 \pi a}} \approx e^{ i k_F x_j } \psi_L ( x_j ) +
e^{ - i k_F x_j } \psi_R ( x_j ) \:\:\: ,
\label{equ3.9}
\eneq
\noindent
with $ x_j = j a$; one gets, then

\[
H_K \approx J \sin ( k_F a )
 \sum_{ | p | \leq \Lambda }  p  [ a_L^\dagger ( p )
a_L ( p ) - a_R^\dagger ( p ) a_R ( p ) ]
\]

\beq
= - i v_F \int_0^L d x \; \biggl[ \psi_L^\dagger ( x )
\frac{ d \psi_L ( x )  }{ d x } -  \psi_R^\dagger ( x )
\frac{ d \psi_R ( x )  }{ d x } \biggr]
\label{equ3.10}
\eneq
\noindent
where the Fermi velocity is given by $v_F =  2 \pi J \sin ( a k_F )$.

Eq.(\ref{equ3.10}) is, of course, the effective low-energy theory of
the hopping Hamiltonian $H_K$;  the cutoff $\Lambda$ will be
specified later.

The dynamics of  the fermionic fields $\psi_L$ and $\psi_R$ in the
Heisenberg representation, is described by

\[
\psi_L ( x , t ) = \psi_L ( x - v_F t ) = \frac{1}{\sqrt{L}}
\sum_p e^{ i p ( x - v_F t ) } \psi_L ( p )
\]
\beq
\psi_R ( x , t ) = \psi_R ( x + v_F t ) = \frac{1}{\sqrt{L}}
\sum_p e^{ i p ( x  + v_F t ) } \psi_R ( p ) \:\:\: ,
\label{equ3.11}
\eneq
\noindent
and the equal time anticommutation relations  are given by

\beq
\{ \psi_L ( p ) , \psi_L^\dagger  ( p^{'} ) \} = \delta_{  p , p^{'}}
\; ; \;
\{ \psi_R ( p ) , \psi_R^\dagger  ( p^{'} ) \} = \delta_{  p , p^{'}}
\; ; \; \{ \psi_R ( p ) , \psi_L^\dagger  ( p^{'} ) \} = 0 \;\;\; .
\label{equ3.12}
\eneq
\noindent

Since $\psi_L^\dagger ( p )$ with $p > 0 $ creates positive energy
left-handed states, while  $\psi_R^\dagger ( p )$ creates positive energy
right-handed states if $ p < 0 $,  the ``Fermi sea'' fermionic ground state
is defined as

\beq
| { \rm FS} \rangle = \prod_{ p < 0 } [ \psi_L^\dagger ( p ) \psi_R^\dagger
( - p ) ] | 0 \rangle \:\:\: ( \psi_L ( p ) | 0 \rangle = \psi_R ( p )
| 0 \rangle = 0  )\;\; .
\label{equ3.13}
\eneq
\noindent

Thus, by choosing $\Lambda$ = $ 1 / ( 4 a )$, one gets:

\beq
\langle {\rm FS} |
2 \pi  a [ e^{ - 2 i k_F x_j } \psi_L^\dagger ( x_j ) \psi_R ( x_j )
+
 e^{  2 i k_F x_j } \psi_R^\dagger ( x_j ) \psi_L ( x_j ) ]
| {\rm FS} \rangle = 0
\label{equ3.14}
\eneq
\noindent

and:

\beq
\langle {\rm FS} | a [ \psi_L^\dagger ( x_j ) \psi_L ( x_j ) +
\psi_R^\dagger ( x_j ) \psi_R ( x_j ) ] | {\rm FS} \rangle
= \frac{1}{2} \:\:\: ;
\label{equ3.15}
\eneq
thus,  $S_j^z  $ is normal ordered respect to $| {\rm FS} \rangle $, i.e.,

\[
 S^z_j  = 2 \pi a [ : \psi_L^\dagger ( x_j ) \psi_L ( x_j ) :
 + : \psi_R^\dagger ( x_j ) \psi_R ( x_j ) : ] +
\]

\beq
2 \pi a  [  : \psi_L^\dagger ( x_j ) \psi_R ( x_j ) : e^{ - 2 i k_F x_j }
 + : \psi_R^\dagger ( x_j ) \psi_L ( x_j ): e^{  2 i k_F x_j }  ]
\label{equ3.16}
\eneq
\noindent
where $::$ denotes normal ordering.

Using fermionic coordinates, one should now evaluate  the Ising-N\'eel
interaction $H_P$ as

\[
H_P \equiv \Delta \sum_{ j = 1}^N S_j^z  S_{ j + 1}^z \equiv
 \Delta \sum_{j=1}^{L / a} ( a_j^\dagger a_j - \frac{1}{2} )
( a_{j + 1}^\dagger a_{j + 1} - \frac{1}{2} )
\approx
\]

\[
( 4 \pi^2 a \Delta ) \int_0^L d x_j \; \{ [ : \psi_L^\dagger ( x_j )
\psi_L ( x_j ) : + : \psi_R^\dagger ( x_j ) \psi_R ( x_j ) :
\]

\[
 + e^{ - 2 i k_F x_j} \psi_L^\dagger ( x_j ) \psi_R ( x_j ) +
e^{  2 i k_F x_j} \psi_R^\dagger ( x_j ) \psi_L ( x_j ) ]
\times
\]

\[
[   : \psi_L^\dagger ( x_{ j + 1} )
\psi_L ( x_{ j + 1}  ) : + : \psi_R^\dagger ( x_{ j + 1} )
\psi_R ( x_{ j + 1} ) : +
\]

\beq
 e^{ - 2 i k_F x_{ j + 1 } }
\psi_L^\dagger ( x_{ j + 1}  ) \psi_R ( x_{ j + 1}  )
+ e^{  2 i k_F x_{ j + 1} } \psi_R^\dagger ( x_{ j + 1}  )
\psi_L ( x_{ j + 1} ) ] \} = H_P^{(1)} + H_P^{(2)} \;\;\; ,
\label{equ3.17}
\eneq
\noindent

where

\[
H_P^{(1)} = ( 4 \pi^2 a \Delta ) \int_0^L d x_j \: [ : \psi_L^\dagger
( x_j ) \psi_L ( x_j ): : \psi_L^\dagger ( x_{j + 1} ) \psi_L ( x_{j + 1} ) :
\]

\[
+  : \psi_R^\dagger ( x_j )
\psi_R ( x_j ): : \psi_R^\dagger ( x_{j + 1} ) \psi_R ( x_{j + 1} ) :
\]

\beq
+  : \psi_L^\dagger ( x_j )
\psi_L ( x_j ): : \psi_R^\dagger ( x_{j + 1} ) \psi_R ( x_{j + 1} ) :
+  : \psi_R^\dagger ( x_j )
\psi_R ( x_j ): : \psi_L^\dagger ( x_{j + 1} ) \psi_L ( x_{j + 1} ) : ]
\:\:\: ,
\label{equ3.18}
\eneq
\noindent
and

\[
H_P^{  ( 2 )} = ( 4 \pi^2 a \Delta )
\int_0^L d x_j \: [ \psi_L^\dagger ( x_j)
\psi_R ( x_j ) e^{ - 2 i k_F x_j}
 + \psi_R^\dagger ( x_j ) \psi_L ( x_j )
e^{ 2 i k_F x_j } ]
\times
\]

\beq
[  \psi_L^\dagger ( x_{ j + 1} )
\psi_R ( x_{ j + 1}  ) e^{ - 2 i k_F x_{ j + 1} }
+ \psi_R^\dagger
( x_{ j + 1} ) \psi_L ( x_{ j + 1} )  e^{ 2 i k_F x_{ j + 1 } }]
\:\:\: .
\label{equ3.19}
\eneq
\noindent
While evaluating $H_P^{(1)}$ is rather straightforward, since it contains
only normal-ordered fermionic left- and right-  densities, evaluating
$H_P^{(2)}$ is a little bit more involved, due to ``crossed''
$L-R$-interaction. In fact, at any  $k_F$, momentum conservation selects
the pertinent contribution to  Eq.(\ref{equ3.19}), given by

\[
H_P^{( 2 ) } =
( 4 \pi^2 a \Delta ) \int_0^L d x_j  \: [ e^{ 2 i k_F a }
\psi_L^\dagger ( x_j ) \psi_R ( x_j ) \psi_R^\dagger ( x_{ j + 1 } )
\psi_L ( x_{ j + 1 } )  +
\]

\beq
e^{ - 2 i k_F a } \psi_R^\dagger ( x_j ) \psi_L ( x_j ) \psi_L^\dagger (
x_{ j + 1 } ) \psi_R ( x_{ j + 1 } ) ] \:\:\: ,
\label{equ3.20}
\eneq
\noindent
where a possible ``Umklapp'' contribution, arising  when $k_F a \sim \pi / 2$,
has been neglected
\footnote{This is the case, for instance,  of the ``half filled'' fermionic
sea in zero chemical potential}. To normal order $H_P^{(2)}$, one may
rewrite it as

\[
H_P^{(2)} =
- ( 4 \pi^2 a \Delta )
e^{ 2 i k_F a } \int_0^L d x \: \biggl[ : \psi_L^\dagger ( x )
\psi_L ( x + a ) : + \frac{ i }{ 2 \pi a} \biggr]
\biggl[:
\psi_R^\dagger ( x + a ) \psi_R ( x ) : + \frac{ i }{ 2 \pi a} \biggr]
\]

\beq
- ( 4 \pi^2 a \Delta ) e^{ - 2 i k_F a } \int_0^L d x \: \biggl[ :
\psi_L^\dagger ( x + a ) \psi_L ( x  ) : - \frac{ i }{ 2 \pi a} \biggr]
\biggl[: \psi_R^\dagger ( x ) \psi_R ( x + a) : - \frac{ i }{ 2 \pi a}
\biggr] \:\:\: ,
\label{equ3.20bis}
\eneq
\noindent
which, for $a \to 0$, becomes

\[
H_P^{ (2)}  = - 2 ( 4 \pi^2 a \Delta ) \cos ( 2 k_F a )
\int_0^L d x \:
: \psi_L^\dagger ( x ) \psi_L ( x ) : : \psi_R^\dagger ( x ) \psi_R ( x ) :
\]

\[
+   4 \pi \Delta \sin ( 2 k_F a)
\int_0^L d x \: [ : \psi_L^\dagger ( x ) \psi_L ( x ) : +
: \psi_R^\dagger ( x ) \psi_R ( x ) : ]
\]

\beq
-  i  4 \pi a \Delta \cos ( k_F a )
\int_0^L d x \:
\biggl[ \psi_L^\dagger ( x ) \frac{ d \psi_L ( x ) }{ d x } -
\psi_R^\dagger ( x ) \frac{ d \psi_R ( x ) }{ d x } \biggr]
\:\:\: .
\label{equ3.21}
\eneq
\noindent

The various terms in Eq.(\ref{equ3.21}) may be interpreted as follows:

\begin{itemize}

\item A shift in the chemical potential:

\[
 4 \pi \Delta \sin ( 2 k_F a)
\int_0^L d x \: [ : \psi_L^\dagger ( x ) \psi_L ( x ) : +
: \psi_R^\dagger ( x ) \psi_R ( x ) : ] \:\:\: ,
\]

which is accounted for by simply redefining $k_F$ through the equation

\beq
- ( a J)  \cos ( k_F a ) + 2  \Delta \sin ( 2 k_F a)
= H \:\:\: ;
\label{equ3.22}
\eneq
\noindent

\item A $L-R$ interaction term:

\[
- 2 ( 4 \pi^2 a \Delta ) \cos ( 2 k_F a )
\int_0^L d x \:
: \psi_L^\dagger ( x ) \psi_L ( x ) : : \psi_R^\dagger ( x ) \psi_R ( x ) :
\:\:\: ,
\]

that adds up to a similar term coming from $H_P^{ ( 1 )}$, giving

\beq
2 ( 4 \pi^2 a \Delta ) [ 1 - \cos ( 2 k_F a ) ]
\int_0^L d x \: : \psi_L^\dagger ( x ) \psi_L ( x ) :
: \psi_R^\dagger ( x ) \psi_R ( x ) : \:\:\: ;
\label{equ3.23}
\eneq
\noindent

\item A  renormalization of the Fermi velocity given by

\beq
-  i 4 \pi a \Delta \cos ( k_F a )  \int_0^L d x \:
\biggl[ \psi_L^\dagger ( x ) \frac{ d \psi_L ( x ) }{ d x } -
\psi_R^\dagger ( x ) \frac{ d \psi_R ( x ) }{ d x } \biggr]
\:\:\: .
\label{equ3.24}
\eneq
\noindent

\end{itemize}

Using the well-known bosonization rules (\ref{appe7}),
the fermionic Hamiltonian $H^f_{\rm Eff}$ may be written in bosonic
coordinates as

\beq
H^b = \frac{v_F + g_2}{4 \pi} \int_0^L d x \:
\left[ \left( \frac{\partial \phi_L}{
\partial x} \right)^2 + \left( \frac{\partial\phi_R}{\partial x}\right)^2
\right]
 +  2 \frac{g_4}{ 4 \pi } \int_0^L d x \:
\left[ \frac{\partial \phi_L}{
\partial x} \frac{\partial\phi_R}{\partial x}\right] \:\:\: ,
\label{equ3.30}
\eneq
\noindent
where $g_2 = g_4 = 4 \pi ( a \Delta ) [ 1 - \cos ( 2 k_F a ) ]$.

One may readily see that $H^b$ corresponds to the Hamiltonian
for a free, massless, real bosonic field $\Phi$ in 1+1 dimensions, which is
described by the Hamiltonian

\beq
H [ \Pi , \Phi ] =  \frac{v}{ 4 \pi} \int_0^L d x \:
\biggl[ \frac{ 4 \pi^2}{g} \Pi^2 +
g \biggl( \frac{ \partial \Phi}{ \partial x} \biggr)^2 \biggr]
\label{equ3.32}
\eneq
\noindent
where the momentum conjugate to $\Phi$ is $\Pi = ( 2 \pi / g ) \frac{ \partial
\Phi}{ \partial t}$.

Upon defining  two independent chiral fields, $\phi_L^g$ and $\phi_R^g$, as

\[
\frac{\partial \phi_L^g ( x - v t )}{ \partial x}
= \frac{1}{\sqrt{2}} \biggl[ \frac{2 \pi}{\sqrt{g}}  \Pi + \sqrt{g}
\frac{\partial \Phi}{\partial x} \biggr]
\]

\beq
\frac{ \partial \phi_R^g ( x + v t )}{ \partial x}
= \frac{1}{\sqrt{2}} \biggl[ - \frac{2 \pi}{\sqrt{g}}  \Pi + \sqrt{g}
\frac{\partial \Phi}{\partial x} \biggr] \:\:\: ,
\label{equ3.33}
\eneq
\noindent
one immediately sees that

\beq
H [ \Pi , \Phi ] \to H [ \phi_L^g , \phi_R^g ] =
\frac{v}{ 4 \pi} \int_0^L d x \: \left[ \left( \frac{ \partial \phi_L^g}{
\partial x} \right)^2 + \left( \frac{ \partial \phi_R^g}{ \partial x}
\right)^2 \right] \:\:\: ,
\label{equ3.34}
\eneq
\noindent
which, when expressed in terms of  $ \phi_L^{ g = 1}$ and $\phi_R^{g=1}$,
yields Eq.(\ref{equ3.30}), provided that

\beq
v = \sqrt{ ( v_F + g_2 )^2 - g_4^2 } \;\;\; ; \;
g = \sqrt{\frac{ v_F + g_2 + g_4}{v_F + g_2 - g_4} } \:\:\: .
\label{equ3.36}
\eneq
\noindent

Thus, the correlation functions of all the operators depending on
$\phi_L$ and $\phi_R$ may be evaluated by the replacements

\beq
\phi_L - \phi_R = \sqrt{g} [ \phi_L^g - \phi_R^g ] \;\;\; , \;\;
\phi_L + \phi_R = \sqrt{\frac{1}{g}} [ \phi_L^g + \phi_R^g ] \;\;\; ,
\label{equ3.37}
\eneq
\noindent
with $\phi_L^g$, $\phi_R^g$ free, chiral bosonic fields.

\section{Phase diagram of the JJ-chain}

In Ref.\cite{glar}, it has been evidenced that the phases allowed to a
JJ-chain are:

\begin{enumerate}

\item A ``Mott insulating'' (MI) phase;

\item A ``band isulating'' (BI)  phase;

\item A Repulsive Tomonaga-Luttinger phase (RTL);

\item A Superconducting (S), attractive Tomonaga-Luttinger phase.

\end{enumerate}

Here, we shall determine the range of the JJ-chain parameters
associated to each allowed phase in the $V_g-J$-plane and, using the
TLL approach, we shall provide a careful derivation of the phase
boundaries; of course, our results crucially depend on the
approximations made in Section 2. Our subsequent analysis is based
on the bosonic, low-energy effective Hamiltonian given in
Eq.(\ref{equ3.30}).

To analyze the onset of the MI phase, one has  to include also the Umklapp
term $H_P^{u}$ in Eq.(\ref{equ3.19}), given by

\beq
H_P^{u} \approx -
( a \Delta ) \int_0^L d x_j \: [ \psi^\dagger_L ( x_j )
\psi_L^\dagger ( x_{j + 1 } ) \psi_R ( x_j ) \psi_R ( x_{j + 1 } )
 +  \psi^\dagger_R ( x_j )
\psi_R^\dagger ( x_{j + 1 } ) \psi_L ( x_j ) \psi_L ( x_{j + 1 } ) ]
\:\:\: ,
\label{equ31.1}
\eneq
\noindent
whose bosonized version yields

\beq
H_P^{u} =
- \left( 2 \frac{g_U}{L^2} \right) \int_0^L d x
\: : \cos [ 2 \sqrt{2} \Phi ( x ) ] : \:\: \: ; \:
 g_U =  a \Delta
  \biggl( \frac{2 \pi a}{L} \biggr)^\frac{4}{g} \:\:\: .
\label{equ31.2}
\eneq
\noindent

Eq.(\ref{equ31.2}) and Eq.(\ref{equ3.32}) yield the Hamiltonian for a
1+1-dimensional Sine-Gordon model, whose phase structure,
as a function of the parameters $g$ and $g_U$ has been extensively
studied \cite{sinegordon}. There are two distinct regimes: if  $g < 2$,
the interaction is irrelevant and the theory is perturbative in $g_U$,
while, if $g > 2$, the interaction  is relevant. In the
thermodynamic limit ($L \to \infty$) the system flows towards a
strongly-coupled regime, where the Umklapp interaction is responsible
for the creation of a gap in the excitation spectrum and for the onset of
long range Ising-N\'eel order \cite{umklapp}. In the language of the JJ-chain,
this corresponds to a checkboard charge ordered state with the charge
at each grain either equal to $n$, or to $n + 1$: this is the MI-phase.

The MI charge-ordered region in the $V_g-J$ plane may be identified with
the condition $ g > 2$, which reads

\beq
4 \sin ( a k_F ) \left( E_Z - \frac{3}{16} \frac{J^2}{E_C } \right)
> \frac{3}{2} J
\;\:\: .
\label{equ31.6}
\eneq
\noindent

As $J =0$, Eq.(\ref{equ31.6}) is satisfied by any value of $k_F$ (provided
that there is a real solution to Eq.(\ref{equ3.22})) When there is no
real solutions to Eq.(\ref{equ3.22}), that is, for large enough $|H|$, the
chain undergoes a phase transition towards the BI phase. This shows that,
for $J = 0$, the only possible phases are either the BI phase,
or the MI charge-ordered phase.

To see how the transition towards the BI phase extends
for $J>0$, one may start again from Eq.(\ref{equ3.22}), describing
the booundary of the BI phase.  If $ H > 0$,
there are no real solutions of Eq.(\ref{equ3.22}) for

\beq
H - 2 \Delta > J \;\; \Rightarrow \;\; H - 2 E_Z - J + \frac{3}{8} \frac{J^2}{
E_C} > 0 \;\;\; .
\label{equ31.4}
\eneq
\noindent

As $H < 0$, on the other hand, there are no real solutions if

\beq
H + 2 \Delta < - J \;\; \Rightarrow \;\;
H + 2 E_Z + J - \frac{3}{8} \frac{J^2}{
E_C} < 0 \;\;\; .
\label{equ31.5}
\eneq
\noindent
Eqs.(\ref{equ31.4},\ref{equ31.5}) define two regions in the phase diagram
corresponding to BI phases, since the density of charge-carrying states at
the Fermi surface is 0.

Furthermore, as $J > 0$, Eq.(\ref{equ31.6}) admits real solutions only if

\beq \frac{ \frac{3}{8} J }{ \left( E_Z - \frac{3}{16}
\frac{J^2}{E_C } \right) } < 1 \:\:\: \Rightarrow \:\: J < \sqrt{
E_C^2 + \frac{16}{3} E_Z E_C } - E_C \:\:\: , \label{equ31.7} \eneq
\noindent which implies that the MI phase closes when $J =  \sqrt{
E_C^2 + \frac{16}{3} E_Z E_C } - E_C$. Since $\Delta$ changes sign
for $ J = J^* =  \sqrt{ \frac{16}{3} E_Z E_C }$, one finds that, as
the MI phase closes, the Tomonaga-Luttinger liquid interaction is
still repulsive (that is, $g > 1$).

The phase where, instead, the Tomonaga-Luttinger liquid is
attractive (which is a necessary condition, to achieve
superconducting correlations in the 1-d system) takes place for $ J
> J^*$, that is, for $g<1$.

\begin{figure}
\includegraphics*[width=1.0\linewidth]{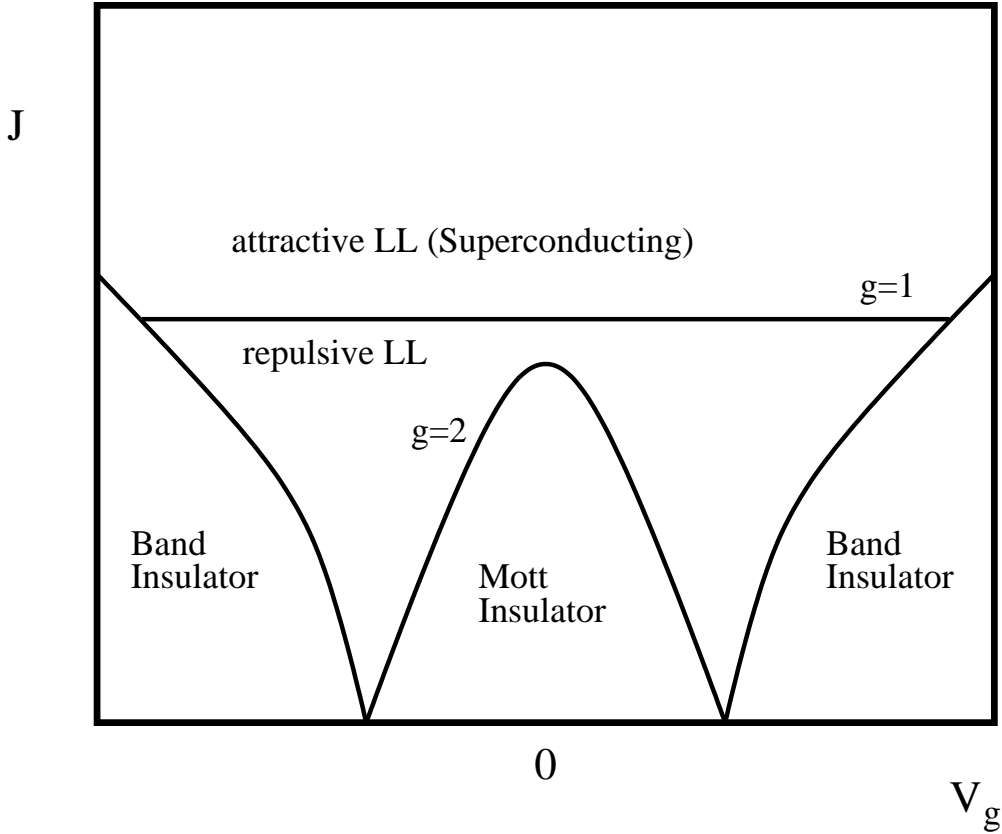}
\caption{Sketch of the phase diagram of the JJ-chain in the $V_g-J$ plane
derived within TLL-approach, as discussed in Section IV.}
\label{figone}
\end{figure}

In Fig.\ref{figone} we plot the phase diagram obtained using the
TLL-approach. We observe that, due to the renormalization of the
Fermi velocity, the line corresponding to $g=1$ is a straight
horizontal line: thus, as long as the TLL- description of the
JJ-chain holds, one cannot push the system across this line by
acting on the gate voltage $V_g$. We expect that this behavior is a
byproduct of the approximations introduced in Section 2; higher
order contributions to $H_{\rm eff}$ should strongly modify the line
corresponding to $g=1$.

\section{Two-boundary Sine-Gordon-model description of a finite JJ-chain}

In the following,  we shall consider a one-dimensional JJ-chain with
a weak link (i.e., a junction with a different nominal value of the
Josephson coupling, $E_W$) at its center, whose position is set at
$x = 0$, and ending with two bulk superconductors, whose phase
difference is held fixed at $\varphi$  (i.e.,  $\varphi_R = -
\varphi_L = \varphi/2$). Using the bosonization method, in this
Section we show that this finite JJ-chain is pertinently described
by a two-boundary Sine-Gordon model \cite{2bsg}.

Upon introducing JW fermions on both sides of the weak link, one gets

\[
S_{ L / a  , > }^+ =
 [ e^{ i \pi \sum_{ l=1  }^{ L / a - 1} a_l^\dagger a_l }]  a_{ L / a}^\dagger
=  e^{ i \varphi / 2 }
\]

\beq
S_{ - L / a , < }^+  = [ e^{ i \pi \sum_{ l = -1  }^{ - L / a - 1} a_l^\dagger
a_l } ] a_{ - L / a}^\dagger =  e^{- i \varphi / 2 }
\label{equ4.3}
\eneq
\noindent
where the labels  $_>$ ($_<$) refer to observables at the right (left)-hand
side of the weak link.

Using the long wavelength approximation, the fermionic string in the
exponential of Eqs.(\ref{equ4.3}), is easily evaluated as:

\[
\pi \sum_{ l = 1 }^{ L / a - 1 } a_l^\dagger a_l
= \frac{\pi L}{ 2 a } + \int_{ 0^-} ^{ L - a } d x_l \:
[ : \psi_{ L , > }^\dagger ( x_l ) \psi_{ L , > }
( x_l ) :
 +  : \psi_{ R , > }^\dagger ( x_l ) \psi_{ R , > } ( x_l ) : ]
=
\]

\beq
 \frac{\pi L}{ 2 a } + \frac{1}{2} [ \phi_{ L , > }  ( L - a ) + \phi_{ R , > }  ( L - a ) ]
\label{equ4.5} \:\:\: ,
\eneq
\noindent
which, in turn, implies:

\beq
S^+_{ L /a , >  }  \approx
 \left( \frac{ 2 \pi a}{L}
\right)^\frac{1}{2}
: e^{ \frac{3 i }{2}  \phi_{ L , > }  ( L ) } : e^{ \frac{i}{2}
\phi_{ R , > }  ( L ) } :
+ \left( \frac{ L }{ 2 \pi a} \right)^\frac{1}{2} : e^{ \frac{i}{2}
\phi_{ L , > } ( L ) } : : e^{ - \frac{i}{2} \phi_{ R , > }  ( L ) } :
\:\:\: .
\label{equ4.6}
\eneq
\noindent

and, by keeping only the leading contributions to Eq.(\ref{equ4.6})
in the cutoff $a$,  as $a \to 0$, one gets, for $x > 0$,

\beq
S^+_{ L/a , > }  \approx
\left( \frac{ L }{ 2 \pi a} \right)^\frac{1}{2} : e^{ \frac{i}{2}
\phi_{ L , > }  ( L ) }: : e^{ - \frac{i}{2} \phi_{ R , >} ( L ) } :
=  e^{ \frac{i}{ 2 } [ \phi_{ L , > } ( L ) -
\phi_{R , > } ( L ) ] } \:\:\: .
\label{equ4.7}
\eneq
\noindent

Similarly, for $x<0$, one obtains

\beq
 S^+_{ - L/a , < } =  e^{ \frac{i}{ 2 } [ \phi_{ L , < } ( -L ) -
\phi_{R , < } ( -L ) ] } \:\:\: .
\label{equ4.8}
\eneq
\noindent

The boundary condition at $x = 0$ is, instead, dynamical, since it depends on
the strength of the weak link, $E_W$. In terms of the spin variables, the
weak link interaction may be represented as a  pointwise contact Hamiltonian
given by

\beq
H_W = \frac{E_W}{2} [ S_{ 0 , <}^+ S_{0 , > }^- + S_{ 0 , > }^+
S_{ 0 , < }^- ] \:\:\: .
\label{equ4.10}
\eneq
\noindent

Taking into account that $S_0^z =  \frac{1}{2} [
a_0^\dagger a_0 + a_0 a_0^\dagger ] = \frac{a}{L} [ : \psi^\dagger_L ( 0 )
\psi_L ( 0 ) + \psi_R^\dagger ( 0 ) \psi_R ( 0 ) :]$ and the requirement that
$S_0^+ S_0^- - S_0^- S_0^+ = 2 S_0^z$, for any value of $g$,
the operators $S_0^+$ and $S_0^-$ are realized as:

\[
S_0^+ =
\frac{a}{L} \left( \frac{2 \pi a}{L} \right)^\frac{g }{2}  : e^{  \frac{i}{2}
[ \phi_L ( 0 ) - \phi_R ( 0 ) ] } :
\]

\beq
S_0^- =
\frac{a}{L} \left( \frac{2 \pi a}{L} \right)^\frac{g }{2}  : e^{ -
 \frac{i}{2} [ \phi_L ( 0 ) - \phi_R ( 0 ) ] } : \:\:\: .
\label{equ4.11}
\eneq
\noindent

>From Eqs.(\ref{equ4.11}), the dependence of $H_W$ on the bosonic coordinates
is given by:

\beq
H_W = \frac{a E_W}{2 L} \left( \frac{2 \pi a}{L} \right)^g
\biggl\{ : \exp \biggl[  \frac{i}{2} [ - \phi_{ L , < } ( 0 ) +
\phi_{R , < } ( 0 ) - \phi_{ L , > } ( 0 ) + \phi_{ R , > } ( 0 )  \biggr]
:
+ {\rm h.c.}  \biggr\}  \:\:\: .
\label{equ4.12}
\eneq
\noindent

Using Eq.(\ref{equ3.37}), one immediately sees that the boundary
interaction Hamiltonian at $x=0$ takes the form

\beq
H_W = \frac{ a E_W}{ 2 L } \left( \frac{ 2 \pi a}{L} \right)^g
 [ : e^{  i \sqrt{\frac{g}{2} }
[ \phi_{ L , + } ( 0  ) - \phi_{ R , +  } ( 0 ) ] } :
+  : e^{ -  i \sqrt{\frac{g}{2} }
[ \phi_{ L , + } ( 0  ) - \phi_{ R , +  } ( 0 ) ] } : ]
\label{equ4.14}
\eneq
\noindent

where

\[
\frac{ \partial \phi_{L , +} ( x - v t )}{ \partial x}  = \frac{1}{ \sqrt{2}}
\left[ \frac{2 \pi}{ \sqrt{g}} \Pi_+ + \sqrt{g} \frac{ \partial \Phi_+}{
\partial x} \right]
\]

\beq
\frac{ \partial \phi_{R , +}  ( x + v t )}{ \partial x}  = \frac{1}{ \sqrt{2}}
\left[ - \frac{2 \pi}{ \sqrt{g}} \Pi_+ + \sqrt{g} \frac{ \partial \Phi_+}{
\partial x} \right] \:\:\: ,
\label{equ4.15}
\eneq
\noindent
with $\Phi_\pm ( x , t ) = \frac{1}{ \sqrt{2}} [ \Phi_> ( x , t ) \pm
\Phi_< ( - x , t ) ]$.

The boundary conditions at the bulk superconductors may
be written as:

\beq
\sqrt{g} [ \phi^g_{ L , + } (  L , t ) - \phi^g_{ R , + } (  L , t ) ] =
\varphi \: \: ({\rm mod} \: 2 \pi k ) \:\:\: .
\label{equ4.9bis}
\eneq
\noindent
By inspection of Eqs.(\ref{equ4.14}), one
sees that the field $\Phi_- ( x , t )$ fully decouples from the weak
link dynamics. Furthermore, its boundary condition  is
$\Phi_- ( L , t ) = 0 $ $ \forall t$, thus implying that $\Phi_-$ is
insensitive to variations in the phase difference between the bulk
superconductors.
As a result, the field $\Phi_-$ does not contribute to the dynamics of the
JJ-network.  Using   Eq.(\ref{equ4.14}), one gets that the pertinent effective
Hamiltonian $H_{JJ}$ is given by

\beq
H_{ \rm JJ} = \frac{v}{ 4 \pi} \int_0^L d x \: \left[ \left(
\frac{ \partial \phi_L}{
\partial x} \right)^2 + \left( \frac{ \partial \phi_R}{ \partial x} \right)^2
\right]
 - \frac{ a E_W}{L} \left( \frac{ 2 \pi a}{L} \right)^g
:\cos \left[ \sqrt{\frac{g}{2} } ( \phi_L ( 0 ) - \phi_R (
0 ) ) \right] :
\label{equ4.16}
\eneq
\noindent
with $ ( \phi_{L , +} , \phi_{ R , + } ) \to ( \phi_L , \phi_R )$,
while the pertinent boundary condition is given by:

\beq
\sqrt{g} [ \phi_L ( L , t  ) - \phi_R ( L , t ) ]  = \varphi \:\:\: ({\rm mod}
\: 2 \pi k ) .
\label{equ4.17}
\eneq
\noindent

The model described by $H_{JJ}$, supplemented with the boundary
condition in Eq.(\ref{equ4.17}), coincides with the two-boundary
Sine-Gordon Hamiltonian introduced in Eq.(\ref{eq1.1}), provided
$\Phi$ in Eq.(\ref{eq1.1}) is identified with $(\phi_L - \phi_R )
/\sqrt{2}$, $\Delta_R$ is sent to $\infty$, and $\Delta_L$ is
identified with $ \frac{ a E_W}{L} \left( \frac{ 2 \pi a}{L}
\right)^g$. Intuitively speaking, while the boundary condition at
$x=L$ is always Dirichlet-like, at $x=0$, since $E_W$ is finite, the
boundary condition is dynamical, and is given by:

\beq \frac{ a E_W }{L} \left( \frac{ 2 \pi a}{L} \right)^g
 \sqrt{g} \sin \left[ \sqrt{ \frac{g}{2} } ( \phi_L ( 0 , t ) -
\phi_R ( 0 , t ) ) \right]
+ v  \frac{ \partial  }{ \partial x} [ \phi_L ( 0 , t ) - \phi_R ( 0 , t ) ]
= 0 \:\:\: .
\label{equ4.18}
\eneq
\noindent
In particular, for small $E_W$, Eq.(\ref{equ4.18}) provides Neumann-like
boundary conditions for $\phi_L - \phi_R$ at $x = 0$, while it
provides Dirichlet-like boundary conditions for large values of $E_W$.

For $g<1$, the boundary term is a relevant operator and one may use
the Renormalizion Group to describe how the renormalization of
$\bar{E}_W$ affects the ground-state energy: as we shall evidence in
section 6, there is a renormalization group invariant length scale
$L^*$ such that, for $L \geq L^*$ the JJ-chain behaves
nonperturbatively in $\bar{E}_W$ while, for $L < L^*$, it behaves
perturbatively. At variance, for $g > 1$,  the JJ-chain  is always
perturbative in $\bar{E}_W$, since the boundary term is now an
irrelevant operator. For $g=1$, the boundary term is marginal and
the bulk system is fully described by a pair of noninteracting
chiral fermions and the partition function may be computed exactly
\cite{2bsg}.

To summarize, for $g>1$ $\bar{E}_W$ always flows to 0, while, for
$g<1$, there is a characteristic " healing" length $L^* = L \left(
\frac{ J }{ \bar{E}_W ( \Lambda = 1 ) } \right)^\frac{1}{1 - g }$,
separating a small-$\bar{E}_W$ perturbative regime from the
nonperturbative one $\bar{E}_W \geq 1$. Similar features are
exhibited by a superconducting loop closed by a Josephson junction
of strength $E_J$, when $E_J$ is regarded as an effective coupling
strength \cite{sawtooth}.

In the next Section, we shall prove that the behavior of the DC
Josephson current as a function of $\varphi$ depends crucially on
whether $\bar{E}_W$ flows to zero, or to large values. Namely, we
shall find that, when   $\bar{E}_W$ flows to zero, the DC Josephson
current has a sinusoidal behavior, while, when $\bar{E}_W \sim 1$,
one gets the sawtooth behavior.

\section{Josephon current across the JJ-network with a weak link}

In this Section, we shall compute the functional dependence of the
DC Josephson current on the phase difference between the bulk superconductors,
by evaluating  the zero temperature canonical
partition function ${\it Z} [ E_W ]$, from which the Josephson current
may be evaluated as

\beq
I ( \varphi ) = \frac{2e}{c}
\frac{ \partial E_{JJ} [ \varphi ] }{ \partial \varphi}
\label{equa5.1}
\eneq
\noindent
with
\beq
E_{JJ} [ \varphi ] = - \lim_{\beta \to \infty} \frac{1}{\beta}
\ln \left[ \frac{ {\it Z} [ E_W ]}{ {\it Z} [ 0 ] } \right]
\label{equa5.2}
\eneq

The partition function of the two-boundary Sine-Gordon model has
been exactly computed for particular values of $g$ \cite{2bsg}; due
to our interest in providing an estimate of the Josephson current
fon any value of $g$, we resort to an approximate computation based
on the Coulomb Gas Renormalization Group scheme \cite{ayh},
described in detail in Appendix B. Our analysis shows that, while
for Neumann boundary conditions ($E_W \to 0$), one gets $ I (
\varphi ) \propto \sin ( \varphi )$, for Dirichlet boundary
conditions ($E_W \geq 1$) one gets a sawtooth dependence of $I (
\varphi )$ on $\varphi$.

\vspace{0.5cm}

{\it Case A: $E_W \to 0$}

 \vspace{0.5cm}

 As $E_W \sim 0$, Eq.(\ref{equ4.18}) provides Neumann-like
boundary condition at $x=0$,

\beq
\frac{ \partial}{ \partial x} [ \phi_L ( 0 -  v t ) -
\phi_R ( 0 + v t) ] = 0 \:\:\: ,
\label{equ5.1}
\eneq
\noindent

together with the Dirichlet-like boundary
condition at $x=L$:

\[
\phi_L ( L -  v t ) - \phi_R ( L  +  v t ) = \frac{ \varphi}{
\sqrt{g}}  \:\:\: .
\]
\noindent

Using the mode expansion given in Eq.(\ref{appe3}), one sees that
Eq.(\ref{equ5.1}) is satisfied if

\[
q_L - q_R = \frac{ \varphi}{ \sqrt{g}}  \;\; ; \: p_L = p_R \equiv p
\]

\beq \phi_L ( n ) e^{ i k_n L } = -  \phi_R ( - n ) e^{ - i k_n L }
\equiv \alpha ( n ) \;\;\; ; \; ( k_n = \frac{ 2 \pi}{L} \left ( n +
\frac{1}{4} \right) \; , \; n \in Z ) \label{equ5.2} \eneq \noindent

which, in turn, implies

\beq
\phi_L ( 0 -  v t ) - \phi_R ( 0 + v t ) =
\frac{\varphi}{ \sqrt{g}} + \frac{ 4 \pi}{L} \sum_{ n \neq 0 }
\frac{ e^{ - i k_n v t}}{ k_n } \alpha ( n )  \equiv \chi ( t ) \:\:\: .
\label{equ5.3}
\eneq
\noindent

The relevant vertex operators are given by

\beq
: e^{ \pm  i \sqrt{g} \chi ( t ) } : = e^{ \pm i \varphi} e^{
\pm i \sqrt{g} \chi_+ ( t ) } e^{ \pm i \sqrt{g} \chi_- (t ) }
\label{equ5.4}
\eneq
\noindent
with

\beq
\chi_+ ( t ) = \frac{ 4 \pi }{L} \sum_{ n < 0 } \frac{ e^{ - i k_n  v
t }}{ k_n} \alpha (n ) \;\; ; \;
\chi_- ( t ) = \frac{ 4 \pi  }{L} \sum_{ n > 0 }  \frac{ e^{ - i k_n v
t }}{ k_n} \alpha (n )  \;\;\; .
\label{equ5.5}
\eneq
\noindent

The  partition function is given by

\beq
{\it Z }[ E_W ] = {\it Z} [ E_W = 0 ] \left\langle T_\tau
\left[ \exp \left[ \frac{a E_W}{L} \left( \frac{2 \pi a}{L}
\right)^g \int_0^\beta d \tau \: : \cos [ \sqrt{\frac{g}{2}} \chi (
\tau ) ] : \right] \right] \right\rangle
\label{equ5.6}
\eneq
\noindent

where $\tau = i t $, $T_\tau$ is the (imaginary) time-ordered
product, and the brackets $\langle \dots \rangle$ mean that
expectation values should be computed with respect to the ground
state of the Hamiltonian in Eq.(\ref{equ3.34}).

Using the expansion of Eq.(\ref{equ5.6})  in a power series of $E_W$
as

\beq
{\it Z }[ E_W ] = {\it Z} [ E_W = 0 ] \sum_{j=0}^{\infty} \frac{
(\bar{E}_W)^j}{ 2^j j!}
\sum_{\alpha_j = \pm 1} \int_0^\beta \prod_{k=1}^j
d \tau_k \:
\left \langle T_\tau [ \prod_{k = 1}^j : e^{ i \alpha_k \chi (\tau_k ) } : ]
\right\rangle \:\:\:
\label{equ5.7}
\eneq
\noindent
with $\bar{E}_W = \frac{a E_W}{L} \left( \frac{2 \pi a}{L} \right)^g$, and
using the identity

\beq
\theta [ \tau_1 - \tau_2 ] [ \chi_- ( \tau_1 ) , \chi_+ ( \tau_2 ) ] =
4 \ln [ 1 - e^{ - \frac{ 2 \pi v ( \tau_1 - \tau_2 ) }{L}}  ]
\label{equ5.8} \:\:\:,
\eneq

one gets

\[
\frac{ {\it Z }[ E_W ]}{ {\it Z} [ E_W = 0 ] } = \sum_{j=0}^{\infty} \frac{
(\bar{E}_W)^j}{ 2^j }
\sum_{\alpha_j = \pm 1} \int_0^\beta d \tau_1 \:
\int_0^{\tau_1 - a / v} d \tau_2 \ldots \int_0^{\tau_{j-1} - a / v}
d \tau_j  \times
\]

\beq
e^{ i \frac{\varphi}{\sqrt{g}} \sum_{k=1}^j \alpha_k}
\prod_{ u < r = 1}^{ 2 j } [ 1 - e^{ \frac{ - 2 \pi v}{L} | \tau_u - \tau_r |
} ]^{  2 g \alpha_u \alpha_r  }  \:\:\: ,
\label{equ5.10}
\eneq

since Wick's theorem implies \footnote{In Eq.(\ref{equ5.10}), the
cutoff $a$ has been introduced to regularize possible short-distance
divergencies in the argument of the integral. It should be
identified with the lattice step introduced in Eq.(\ref{equ2.1}).}

\beq
\langle T_\tau [ \prod_{ k = 1}^{ 2 j } : e^{ i \sqrt{ \frac{ g }{2} }
\alpha_k \chi ( \tau_k ) } : ] \rangle
= e^{ i \frac{\varphi}{\sqrt{g}} \sum_{k=1}^j \alpha_k}
\prod_{ u < r = 1}^{ 2 j } [ 1 - e^{ \frac{ - 2 \pi v}{L} | \tau_u - \tau_r |
} ]^{  2 g \alpha_u \alpha_r  } \:\:\: .
\label{equ5.9}
\eneq
\noindent

To analyze the short-distance divergences in Eq.(\ref{equ5.10}), one has to
rescale the cutoff $a \to a / \Lambda$, $\Lambda > 1$, in order to
approximate $\prod_{ u < r = 1}^{ 2 j } [ 1 - e^{ \frac{ - 2 \pi v}{L}
| \tau_u - \tau_r |} ]^{  2 g \alpha_u \alpha_r  } $ with
$\prod_{ u < r = 1}^{ 2 j } \left|  \frac{  2 \pi v}{L} ( \tau_u - \tau_r )
 \right|^{  2 g \alpha_u \alpha_r  } $. As a result, one obtains

\[
\int_0^\beta d \tau_1 \:
\int_0^{\tau_1 - a / \Lambda v} d \tau_2 \ldots \int_0^{\tau_{j-1}
- a / \Lambda v}
d \tau_j
 e^{ i \frac{\varphi}{\sqrt{g}} \sum_{k=1}^j \alpha_k}
\prod_{ u < r = 1}^{ 2 j } \left|  \frac{  2 \pi v}{L} ( \tau_u - \tau_r )
\right|^{  2 g \alpha_u \alpha_r  } =
\]

\beq
\Lambda^{ g \sum_{u \neq r} \alpha_u
\alpha_r} \int_0^\beta d \tau_1 \:
\int_0^{\tau_1 - a /  v} d \tau_2 \ldots \int_0^{\tau_{j-1}
- a / \ v}
d \tau_j
 e^{ i \frac{\varphi}{\sqrt{g}} \sum_{k=1}^j \alpha_k}
\prod_{ u < r = 1}^{ 2 j } \left|  \frac{  2 \pi v}{L} ( \tau_u - \tau_r )
\right|^{  2 g \alpha_u \alpha_r  } \:\:\: .
\label{equ5.11}
\eneq
\noindent
At a given order $2j$, and for
$\Lambda \to \infty$, the most diverging contributions come from integrals
containing an equal number of positive and negative $\alpha$'s. Thus,
$\Lambda$-scaling of the integrals appearing in Eq.(\ref{equ5.11}) is
taken into account by means of a multiplicative renormalization of the
effective coupling strength

\beq
\bar{E}_W \to \bar{E}_W ( \Lambda ) = \Lambda^{1-g}  \bar{E}_W ( \Lambda = 1)
\:\:\: .
\label{equ5.12}
\eneq
\noindent

Eq.(\ref{equ5.12}) implies that the boundary interaction at $x=0$ is
irrelevant and the Neumann fixed point is always stable for $g > 1$
(i.e., in the repulsive Tomonaga-Luttinger phase). The RTL phase is
always associated to a stable Neumann fixed point.

To evaluate $ I ( \varphi )$, one may  retain only the first order
terms in the $E_W$-expansion in Eq.(\ref{equ5.7}), getting

\beq
\frac{ {\it Z} [ E_W ] }{ {\it Z} [ 0 ] } \approx
1 - \frac{ \bar{E}_W }{
2} \int_0^\beta d \tau \: \langle
e^{ i \varphi} : e^{ i \sqrt{\frac{g}{2}}
\chi ( \tau ) } : + e^{ - i \varphi} :
e^{ - i \sqrt{\frac{g}{2}} \chi ( \tau ) } : \rangle
\approx e^{ - \beta
( a E_W )   \left( \frac{ 2 \pi a}{L} \right)^{  g} \cos ( \varphi ) }
\:\:\: ,
\label{equ5.13}
\eneq
\noindent
from which  the network energy is derived as

\beq
E_{JJ} [ \varphi ] = - \lim_{ \beta \to \infty } \frac{ 1 }{ \beta }
\ln \frac{ { \it Z  }[ E_W ] }{ { \it Z }  [ 0 ] }
= ( a E_W )   \left( \frac{ 2 \pi a}{L} \right)^{  g} \cos ( \varphi )
\;\;\; .
\label{equ5.14}
\eneq
\noindent
Using Eq.(\ref{equa5.1}) one gets  $I ( \varphi ) \propto \sin ( \varphi )$.

It is comforting to see that Eq.(\ref{equ5.14}) reproduces the pertinent
renormalization of the effective coupling constant given in \cite{glar}.

\vspace{0.5cm}

{\it Case B: $E_W \to 1$}

\vspace{0.5cm}

When analyzing the case in which the effective coupling grows, as the size of
the system goes large, we must consider that our analysis started from
expanding fermionic fields whose band energy is equal to $J$.
Accordingly, the scaling should  stop as $\bar{E}_W \sim J$.
The scale at which this  happens, $\Lambda^*$,  is found by the condition

\beq
\bar{E}_W ( \Lambda = 1 ) ( \Lambda^* )^{1-g}  = J
\label{due}
\eneq
\noindent
This implies that scaling stops as the size of the system becomes of order
of $L^*$, given by

\beq
L^* = L \left( \frac{ J }{ \bar{E}_W ( \Lambda = 1 ) }
\right)^\frac{1}{1 - g }
\label{tre}
\eneq
\noindent

For $L < L^*$, the theory is still perturbative. As $L \sim L^*$,
instead, the system enters the nonperturbative region. In this
limit, the field $\phi_L - \phi_R$ has to satisfy Dirichlet-like
boundary conditions both at $x=L$ and at $x=0$. From
Eq.(\ref{equ4.18}), one gets

\beq \sin [ \sqrt{ \frac{g}{2} }[ \phi_L ( 0 - v t ) - \phi_R ( 0 +
v t ) ] ] = - \frac{1}{\sqrt{g} \bar{E}_W v}
\frac{\partial}{\partial x} [ \phi_L ( 0 - v t  ) - \phi_R ( 0 + v t
) ] \:\:\: ; \label{equ4.19} \eneq \noindent thus, the
Dirichlet-like boundary condition at $x=0$ is

\beq \phi_L ( 0 - v t  ) - \phi_R ( 0 + v t )  = 0 \:\:\:  .
\label{equ4.20} \eneq \noindent Using the mode expansion given in
Eq.(\ref{appe3}), Eq.(\ref{equ4.20}) may be easily satisfied by
setting

\[
q_L = q_R \equiv q \:\: ; \: p_L =  p_R \equiv p
\]

\beq
\phi_L ( n ) e^{ i k_n L } + \phi_R ( - n ) e^{ -  i k_n L} = 0
\:\:\: ,
\label{equ4.21}
\eneq
\noindent
provided that:

\[
k_n = \frac{ 2 \pi n }{L} \:\: ; \: n \in Z \;\; ; \;\; - \sqrt{g} 4
\pi  p =  2 \pi k +  \varphi
\]

\beq
 \phi_L ( n ) = - \phi_R ( - n ) = \alpha ( n ) \:\:\: .
\label{equ4.22}
\eneq
\noindent

The partition function is now given by \beq {\it Z} = {\rm Tr} \exp
\biggl[ - \beta \frac{\pi v}{L} ( p_L^2 + p_R^2 ) - \beta \frac{ 2
\pi v}{L} \sum_{ n > 0 } ( \phi_L ( - n ) \phi_L ( n ) + \phi_R ( n
) \phi_R ( - n ) ) \biggr] \:\:\: , \label{equ4.23} \eneq \noindent
with the pertinent boundary conditions given by Eq.(\ref{equ4.22}).

The trace in Eq.(\ref{equ4.23}) may be factorized into a
contribution from the oscillatory modes, and a contribution from the
zero modes, so that

\beq
{\it Z} = {\it Z}_{\rm osc} {\it Z}_{\rm 0 - modes} \:\:\: ,
\label{equ4.24}
\eneq
\noindent

 with ${\it Z}_{\rm osc} = [ \prod_{
n = 1}^\infty [ 1 - ( e^{ - \beta \frac{ 4 \pi v}{L} } )^n ]
]^{-1}$, and

\beq
{\it Z}_{\rm 0-modes} = \sum_{ k = - \infty}^\infty \exp \left[
- \beta \frac{  v \pi}{2 g L} \left( k - \frac{ \varphi}{ 2 \pi }
\right)^2 \right] \:\:\: . \label{equ4.25} \eneq \noindent

>From Eqs.(\ref{equ4.24},\ref{equ4.25}), one gets $ I ( \varphi )
\propto \varphi$, for $ -  \pi \leq \varphi   \leq \pi$; this yields
the well-known sawtooth behavior \cite{sawtooth}.

The switch to this behavior from the sinusoidal behavior obtained
for $E_W \sim 0$ signals the crossover from a perturbative to a
nonperturbative regime of the chain. It should be observed that, for
$g >1$, the DC Josephson current has always a sinusoidal dependence
on the phase difference between the bulk superconductors since, in
this region, the boundary term is an irrelevant operator and, thus,
the chain's behavior is always bulk-dominated.

\section{Concluding remarks}

Our analysis shows how a two-boundary Sine-Gordon model emerges as a
pertinent effective
description of a finite JJ-chain. Since our analysys
heavily relies on the bosonization method, we expect that boundary
Sine-Gordon models may be useful where the Tomonaga-Luttinger liquid
paradigm is relevant. For instance, a magnetic spin system with a
pertinent impurity at its center and with the spins at its extrema
held fixed, may support a spin current across the chain, with
different behaviors, depending on the boundary conditions around the
impurity. Similarly, one may envisage other applications of our
results to quantum wires \cite{wires} and carbon nanotubes
\cite{nanotubes}.

According to the $g$-theorem \cite{gthe}, the boundary entropy of
the chain should always decrease, as one gets towards the
thermodynamic limit. Thus, for $L \geq L^* \to \infty$, one has

\beq S_D - S_N =  \lim_{\beta \to \infty} [ \lim_{L \to \infty} [
\ln {\it Z}_D - \ln {\it Z}_N ] ] \;\;\; , \label{concl1} \eneq
\noindent where ${\it Z}_{D/N}$ is the partition function computed
with Dirichlet / Neumann boundary conditions, respectively.
Eq.(\ref{concl1}) yields a nonvanishing result only because of the
contribution of the zero modes; namely, from Eq.(\ref{equ4.25}), one
gets that

\beq S_D - S_N = \ln [ \sqrt{g} ] \;\;\; . \label{concl2} \eneq
\noindent

Remarkably, the entropy variation depends on the sign of $\ln
\sqrt{g}$. Thus, as $g > 1$ (i.e., within the RTLL phase) the
Dirichlet boundary entropy is higher than the Neumann boundary
entropy and then, in the thermodynamic limit the system flows from
the Dirichlet to the Neumann fixed point. Conversely, as $g < 1 $
(i.e., within the superconducting region) the Neumann fixed point
carries an  entropy that is higher than the one associated to the
Dirichlet fixed point. Accordingly, the flow now goes from the
Neumann to the Dirichlet fixed point.

The evaluation of the Josephson current from the partition function
of the two-boundary Sine-Gordon model explicitly shows, for $g<1$, a
crossover from a perturbative regime ($\bar{E}_W \sim 0$), in which
the current is a sinusoidal function of the phase difference at the
boundary, $\varphi$, to a nonperturbative regime ($\bar{E}_W / J
\geq 1$), where it exhibits a sawtooth functional dependence on
$\varphi$.

There is a striking, and yet intuitive, similarity between the
finite JJ-chain investigated in this paper, and an rf-SQUID in an
external magnetic field \cite{sawtooth}. For the latter system, a
variation of the flux threaded by a superconducting loop operates
the crossover in the behavior of the Josephson current, as a
function of the applied flux. This might suggest that other very
interesting condensed matter realization of boundary Sine-Gordon
models, may be provided by superconducting loops, interrupted by
two, or more, Josephson junctions \cite{mooji}. Remarkably, using
the results in Section 6,  one immediately sees that the effective
potential of the finite JJ-chain, as a function of the phase
difference at the boundary of the weak link, i.e., $\psi = \langle
\phi_L ( 0 ) - \phi_R ( 0 ) \rangle$, exhibits only one minimum
within the RTLL phase (i.e. $\psi = \varphi$), since $\bar{E}_W \sim 0$,
while it is a two-level quantum system in the superconducting region
($\bar{E}_W / J \geq 1$), provided that $\varphi \sim \pi$.

\vspace*{0.5cm}

 We thank G. Zemba for actively participating to our research efforts
at the very early stages of this work. We acknowledge useful
discussions with G. Delfino, A. de Martino, G. Grignani, H. Saleur,
G. W. Semenoff and insightful correspondence with L. Glazman. The
work has been partly supported by the M.I.U.R. national project "
Josephson Networks for Quantum Coherence and Information" (grant
no.2004027555).

\appendix

\section{Bosonization rules}

In this Appendix the bosonization rules used in this paper are reviewed.

It is a peculiar property of  1+1 dimensional theories that it is possible
to realize chiral fermionic fields in terms of chiral bosonic fields, and
vice versa. If one starts, for instance, from the chiral components of a
free, massless, Klein-Gordon field $\Phi$ in
1+1 dimensions, the equation of motion for $\Phi$ is

\beq
\left[ \frac{\partial^2}{ \partial t^2} - v^2
\frac{\partial^2}{ \partial x^2} \right] \Phi ( x , t ) = 0 \:\:\: ,
\label{appe0}
\eneq
\noindent
where periodic boundary conditions are assumed. $\Phi$ may be written,
as the sum of two chiral fields as

\beq
\Phi ( x , t ) =  \frac{1}{\sqrt{2}} [ \phi_L ( x - v t ) +
\phi_R ( x + v t ) ]
\label{appe01}
\eneq
\noindent
with $\phi_L$ and $\phi_R$ chiral Fubini-Veneziano fields \cite{pgisp},
whose mode-expansion is given by

\beq
\phi_L ( x , t ) = \phi_L ( x - v t ) = q_L - \frac{ 2 \pi}{L} p_L
( x - v_F t ) +
 \frac{ 2 \pi i}{L}
\sum_{k_n} \frac{ e^{ i k_n ( x - v t )}}{ k_n} \phi_L ( n )
\label{appe1}
\eneq
\noindent
and:

\beq
\phi_R ( x , t ) =
\phi_R ( x + v t ) = q_R + \frac{ 2 \pi }{L} p_R ( x + v t )
  + \frac{ 2 \pi i}{L}
\sum_{k_n} \frac{ e^{ i k_n ( x + v t )}}{ k_n} \phi_R ( n )
\:\:\: .
\label{appe2}
\eneq
\noindent
The basic commutation rules are

\beq
[ q_L , p_L ] = [ q_R , p_R ] = i \;\;\;
; \;
[ \phi_L ( n ) , \phi_L ( m ) ] = n \delta_{ n + m , 0 } \:\:\: ; \:
[ \phi_R ( n ) , \phi_R ( m ) ] = - n \delta_{ n + m , 0 }
\:\:\: ,
\label{appe3}
\eneq
\noindent
and $ \{ k_n \}$ is a (discrete)  set of nonzero modes depending on
the boundary conditions imposed on the bosonic fields (for instance, for
periodic boundary conditions, one gets $k_n = \frac{ 2 \pi n}{L}$, $n \in Z$).

Due to the commutation rules in Eq.(\ref{appe3}), the bosonic vacuum
$ | {\rm Bos} \rangle$ is defined by

\[
p_L | {\rm Bos} \rangle = p_R | {\rm Bos} \rangle = 0
\]

\beq
\phi_L ( n ) | {\rm Bos} \rangle = \phi_R ( - n ) | {\rm Bos} \rangle
= 0 \;\; ( n > 0 ) \:\:\: ,
\label{appe4}
\eneq
\noindent
and one may then define  a creation and an
annihilation part for each field, i.e.,

\beq
\phi_L^+ ( x  ) = q_L + \frac{ 2 \pi i }{L} \sum_{k_n < 0 }
\frac{ e^{ i k_n x}}{ k_n} \phi_L ( n ) \:\: ; \:\:
\phi_L^- ( x ) = - \frac{ 2 \pi}{L} x  p_L + \frac{ 2 \pi i }{L}
\sum_{k_n > 0 } \frac{ e^{ i k_n x}}{ k_n} \phi_L ( n )
\label{appe5}
\eneq
\noindent

and

\beq
\phi_R^+ ( x ) = q_R + \frac{ 2 \pi i }{L} \sum_{k_n > 0 }
\frac{ e^{ i k_n x}}{ k_n} \phi_R ( n ) \:\: ; \:\:
\phi_R^- ( x ) =  \frac{ 2 \pi}{L} x  p_R + \frac{ 2 \pi i }{L}
\sum_{k_n < 0 } \frac{ e^{ i k_n x}}{ k_n} \phi_R ( n )
\:\:\: .
\label{appe6}
\eneq
\noindent

>From the commutators given in Eqs.(\ref{appe3}), one sees that the modes of the
operator $\frac{1}{ 2 \pi} \frac{ \partial \phi_L}{ \partial x}$ obey
the same algebra as the modes of the fermionic density operator, normal
ordered with respect to $ |{\rm FS} \rangle$.
Thus, the fermionic bilinear density operator
$:\psi_L^\dagger \psi_L :$ may be identified with the bosonic density
operator, $\frac{1}{ 2 \pi} \frac{ \partial \phi_L}{ \partial x}$,
provided  that $|{\rm Bos} \rangle$ is identified with $ |{\rm FS}\rangle$. The
same identification may be carried for the $R$-modes.
Therefore, one gets a first bosonization rule

\[
: \psi_L^\dagger ( x - v t ) \psi_L ( x - v t ) : \to
\frac{1}{ 2 \pi} \frac{ \partial \phi_L ( x - v t )}{ \partial x}
\]

\beq
: \psi_R^\dagger ( x + v t ) \psi_R ( x + v t ) : \to
\frac{1}{ 2 \pi} \frac{ \partial \phi_R ( x + v t )}{ \partial x}
\:\:\: .
\label{appe7}
\eneq
\noindent

The second rule is obtained if one identifies the chiral fermionic fields
with normal ordered vertex operators of bosonic fields, defined by

\[
: e^{ i \alpha \phi_{L / R }  ( x \mp v t ) } : =
e^{ i \alpha [ \phi_{ L / R }^+
( x \mp v t  ) + \phi_{ L / R}^- ( x \mp v t ) ] }
e^{ - \frac{ \alpha^2 }{2} [ \phi_{L / R} ^+ ( x \mp v t  ) ,
\phi_{ L / R}^- ( x \mp v t  ) ] }
\]

\beq
=  e^{ i \alpha [ \phi_{L / R } ^+ ( x \mp v t  ) +
\phi_{ L / R } ^- ( x \mp v t ) ] }  e^{ \mp \frac{ i \alpha^2 \pi}{L} ( x
\mp v t ) }
\left( \frac{L}{ 2 \pi a} \right)^\frac{\alpha^2}{2} \:\:\: .
\label{appe8}
\eneq
\noindent
The correspondence rules are now given by

\beq
\psi_L^\dagger ( x - v t ) \to
\frac{1}{ \sqrt{L}} : e^{ i \phi_L( x - v t ) } : \:\: ; \:
\psi_R^\dagger ( x + v t ) \to
\frac{1}{ \sqrt{L}} : e^{ -i \phi_R( x + v t ) } : \:\:\: .
\label{appe10}
\eneq
\noindent
To check the consistency of Eq.(\ref{appe10}), one has to consider the
 ``braiding rule'':

\beq
: e^{ i \alpha \phi_{L / R }  ( x ) } :
: e^{ i \beta \phi_{L / R }  ( y ) } : =
e^{ i \pi \alpha \beta } : e^{ i \beta \phi_{L / R }  ( y ) } :
: e^{ i \alpha \phi_{L / R }  ( x ) } : \:\:\: ,
\label{appe9}
\eneq
\noindent
and the vertex-vertex correlators:

\beq
\langle  {\rm Bos} | : e^{ i \alpha \phi_{L / R }  ( x \mp v t ) } :
: e^{ - i \alpha \phi_{L / R }  ( x^{'} \mp v t^{'} ) } :
|  {\rm Bos} \rangle =
 \left[ \left( \frac{\pm i}{2} \right)
\frac{1}{\sin [ \frac{2 \pi}{L}
( x - x^{'} \mp v ( t - t^{'} ))] } \right]^{\alpha^2}
\:\:\: .
\label{appe9ter}
\eneq
\noindent
>From Eqs.(\ref{appe9},\ref{appe9ter}), one derives the basic anticommutators:

\[
\{ \psi_L ( x - v t ) , \psi_L^\dagger ( x^{'} - v t^{'} ) \} =
\delta [ x - x^{'} - v ( t - t^{'} ) ]
\]
\noindent
and
\beq
\{ \psi_R ( x + v t ) , \psi_R^\dagger ( x^{'} + v t^{'} ) \} =
\delta [ x - x^{'} + v ( t - t^{'} ) ]
\label{appefinal}
\eneq
\noindent
The other correlators used in the paper are derived from
Eq.(\ref{appe9ter}) and from Wick's theorem applied to normal ordered
vertices \cite{pgisp}.

\section{Renormalization Group equations for the JJ-chain with a weak
link}
 In this Appendix, the flow of $\bar{E}_W$ is derived within the Coulomb Gas
renormalizion Group approach.

To analyze the Renormalization Group flow for the JJ-chain with a weak link,
one needs to observe that, at  order $ 2 j $,
the  short-distance most diverging contribution to the partition function is
given by:

\[
(\bar{E}_W)^{ 2 j } \left( \frac{ 2 \pi a}{L}
\right)^{ 2 j g } \int_0^\beta d \tau_1 \: \int_0^{ \tau_1 - a / v } d \tau_2
\: \ldots \int_0^{\tau_{2j-1} - a / v} d \tau_{ 2 j }
\]

\beq
\sum_{ \alpha_1 + \ldots + \alpha_{2j} = 0 } \:\:\: \prod_{u < r = 1}^{ 2 j }
[ 1 - e^{ - \frac{ - 2 \pi v}{L} | \tau_u - \tau_r | }
]^{ 2 g \alpha_u \alpha_r}
\label{appe2.1}
\eneq
\noindent
Rescaling the short-distance cutoff as $a \to
a / \Lambda$ and sending $\Lambda \to \infty$ implies the following
renormalization group scaling equation for $\bar{E}_W$:

\beq
\frac{ d \bar{E}_W ( \Lambda ) }{d \ln \Lambda} = ( 1 - g ) \bar{E}_W
( \Lambda )
\:\:\: .
\label{appe2.2}
\eneq
\noindent
>From Eq.(\ref{appe2.2}), one may easily  identify the cutoff scale $\Lambda^*
= \left( \frac{J}{E_W ( \Lambda = 1)} \right)^\frac{1}{1 - g}$, at which
$\bar{E}_W$ becomes $\sim J$. It means  that a system of size $L$,
 with a weak link of nominal strength $E_W = E_W ( \Lambda = 1 )$,
 ``crosses over'' towards the strongly coupled regime,
as its size is increased to $ L^* = \Lambda^* L$ \cite{glar}.

In addition, it has to be noticed that, to remove the cutoff, one needs a
further renormalization, due to ``one-dimensional charge
annihilation processes''. This may be evidenced, for instance,  by applying
Anderson-Yuval-Hamann analysis of a one-dimensional instanton gas \cite{ayh}.

As the cutoff is rescaled from $a $ to $a / \Lambda$, two charges, of
opposite sign, may annihilate with each other, if they were originally
separated by a distance between $a / ( v \Lambda)$ and $a / v$.  As a result,
the integral at order $2j+2$ should  provide an extra contribution to the
integral at order $2j$, which we are now going to calculate.

Upon defining

\beq
T = \frac{ \tau_+ + \tau_- }{2} \:\:\: ; \:\:
\tau = \tau_+ - \tau_-
\label{appe2.4}
\eneq
\noindent
where $\tau_+$ being the coordinate of the $+1$-charge, and
$\tau_-$ the coordinate of the $-1$-charge, the extra
contribution arising to order $2j$, is given by

\[
(\bar{E}_W )^{ 2 j + 2 } \int_{ a/ ( \Lambda v ) }^{ a / v  } \: d \tau \:
\biggl\{ \int_0^\beta d  \tau_1 \: \int_0^{\tau_1 - a / L } d \tau_2
\ldots \int_0^{ \tau_{ 2 j - 1} - a / v} d \tau_{ 2 j }
\times \left[ \int_0^{
\tau_{ 2 j } } d T +  \int_{
\tau_{ 2 j - 2} - a / v}^{ \tau_{ 2 j - 1} - a / v }  d T + \ldots \right]
\biggr\}
\times
\]

\[
\frac{1}{ \left[ \frac{ 2 \pi v \tau}{L} \right]^{ 2 g } }
\prod_{ u < r = 1}^{2 j } [ 1 - e^{ - \frac{ 2 \pi v}{L} | \tau_u - \tau_r | }
]^{ 2 g \alpha_u \alpha_r}
\exp \left[  2 \tau g \sum_{ k = 1}^{ 2 j }
\alpha_k \frac{\partial}{\partial T}
\ln [ 1 - e^{ - \frac{ 2 \pi v}{L} | T - \tau_k | } ] \right]
\]

\[
+ (\bar{E}_W)^{ 2 j + 2 } \int_{ a / ( \Lambda v ) }^{ a / v  } \: d
\tau \: \biggl\{ \int_0^\beta d  \tau_1 \: \int_0^{\tau_1 - a / L }
d \tau_2 \ldots \int_0^{ \tau_{ 2 j - 1} - a / v} d \tau_{ 2 j }
\left[ \int_{ \tau_{2j} - a / v}^{ \tau_{ 2 j - 1} - a / v  } d T +
\int_{ \tau_{ 2 j - 3} - a / v}^{ \tau_{ 2 j - 2} - a / v } d T +
\ldots \right] \biggr\} \times
\]

\beq
\frac{1}{ \left[ \frac{ 2 \pi v \tau}{L} \right]^{ 2 g } }
\prod_{ u < r = 1}^{2 j } [ 1 - e^{ - \frac{ 2 \pi v}{L} | \tau_u - \tau_r | }
]^{ 2 g \alpha_u \alpha_r} \exp \biggl[  2 \tau g \sum_{ k = 1}^{ 2 j }
\alpha_k
\frac{\partial}{\partial T}
\ln [ 1 - e^{ - \frac{ 2 \pi v}{L} | T - \tau_k |} ] \biggr] \:\:\: .
\label{appe2.5}
\eneq
\noindent

If one expands the exponentials as

\beq
\exp \left[  \pm 2 \tau g \sum_{ k = 1}^{ 2 j } \alpha_k
\frac{\partial}{\partial T}
\ln [ 1 - e^{ - \frac{ 2 \pi v}{L} | T - \tau_k |} ] \right]
\approx
1 \pm 2 \tau g \sum_{ k = 1}^{ 2 j } \alpha_k
\frac{\partial}{\partial T}
\ln [ 1 - e^{ - \frac{ 2 \pi v}{L} | T - \tau_k |} ] \:\:\: ,
\label{appe2.6}
\eneq
\noindent

one may derive the renormalization group equation for $g$ \footnote{In field
theory language, such an extra renormalization
is equivalent to a wavefunction renormalization}. The result is \cite{ayh}:

\beq
g \to g + d g = g - \left( \frac{L}{2 \pi v a} \right)^g g [ \bar{E}_W ]^2
d \ln \Lambda \:\:\: .
\label{appe2.7}
\eneq
\noindent
As expected \cite{sinegordon}, the wavefunction renormalization is needed only
for $g \leq 1$.

This completes the renormalization scheme derived within the perturbative
approach for the system in the presence of a weak link.



\end{document}